\documentclass[
    ,final            
  ]
  {aipproc}

\layoutstyle{8x11double}

\begin{document}

\title{On the Infrared Behaviour of Landau Gauge Yang-Mills Theory
 with Differently Charged Scalar Fields}


\author{Reinhard Alkofer}{
  address={Institut f\"ur Physik, Universit\"at Graz, Universit\"atsplatz 5, 
  8010 Graz, Austria}
}

\author{Leonard Fister}{
  address={Institut f\"ur Theoretische Physik, Universit\"at Heidelberg,
Philosophenweg 16, D-69120 Heidelberg;\\
ExtreMe Matter Institute EMMI, GSI. 
 Planckstr. 1, D-64291 Darmstadt;\\
Institut f\"ur Kernphysik, TU
Darmstadt, Schlossgartenstra\ss e 9, D-64289 Darmstadt, Germany}
}

\author{Axel Maas}{
 address={Institut f\"ur Physik, Universit\"at Graz, Universit\"atsplatz 5, 
  8010 Graz, Austria}
}

\author{Veronika Macher}{
 address={Institut f\"ur Physik, Universit\"at Graz, Universit\"atsplatz 5, 
  8010 Graz, Austria}
}

\begin{abstract}
Recently it has been argued that infrared singularities of the quark-gluon
vertex of Landau gauge QCD can confine static quarks via a linear potential. It
is demonstrated that the same mechanism also may confine fundamental scalar
fields. This opens the possibility that within functional approaches static 
confinement is an universal property of the gauge sector even
though it is formally represented in the functional equations of the matter 
sector. The
colour structure of Dyson-Schwinger equations for fundamental and adjoint scalar
fields is determined for the gauge groups SU(N) and G(2) exhibiting interesting
cancelations purely due to colour algebra.
\end{abstract}

\maketitle

\section{Motivation}

QCD Green's functions are unique in the sense that they may provide a look onto
the detailed relation between hadron phenomenology and fundamental properties of
QCD. On the one hand, they  can be used to describe hadrons in terms of glue and
quarks. To this end one employs either the Bethe-Salpeter equations for  mesons
\cite{Fischer:2010} or the Poincar{\'e} covariant Faddeev equation for baryons
\cite{Eichmann:2010}. On the other hand, they provide insight into the
non-perturbative structure of QCD, and hereby most noticeably into Dynamical
Breaking of Chiral Symmetry (D$\chi$SB), the U$_A$(1) anomaly, and confinement.
Hereby it is interesting to note that D$\chi$SB does not only imply the
dynamical generation of quark masses but also of Lorentz-scalar couplings
between quarks and gluons, see \cite{Alkofer:2008tt,Alkofer:2006gz} and
references therein. The U$_A$(1) anomaly can be related to the infrared
behaviour of Green's function \cite{Kogut:1973ab,Alkofer:2008et} and thus to
confinement.

As for the infrared behaviour of Landau gauge QCD Green's functions it has
become evident that both, Functional Renormalization Group and 
Dyson-Schwinger equations, display two different types of solutions.
There is exactly one unique solution with powerlaw infrared behaviour which is
called scaling solution \cite{Fischer:2009tn}. On the other hand, there is an
one-parameter family of solutions with an infrared constant gluon propagator
and infrared-trivial vertex functions, for a discussion see
\cite{Fischer:2008uz} and references therein. A similar situation is known
since quite some time in Coulomb gauge where it occurs when one applies
variational methods \cite{Szczepaniak:2001rg}. In Landau gauge, however, this
has been noticed only recently, see {\it e.g.\/}
\cite{Fischer:2008uz,Aguilar:2008xm,Boucaud:2008ky}.
A possible resolution of this apparent ambiguity might be that even the minimal
Landau gauge needs some additional input to be fixed also non-perturbatively 
\cite{Maas:2009se}. In the following we will assume that at least one gauge
exists which also allows for the scaling solution.

\section{Landau Gauge Yang-Mills Green's functions}

The derivation and main characteristics of the scaling solution is summarized in
a contribution to the Proceedings of the Confinement Conference 2008 
\cite{Alkofer:2008bs} to which the reader is refered for more details. 
The following general infrared behaviour for one-particle irreducible
Green functions with $2n$ external ghost legs and $m$ external
gluon legs is in the scaling solution given as 
\cite{Alkofer:2004it,Huber:2007kc}:
\begin{equation}
\Gamma^{n,m}(p^2) \sim (p^2)^{(n-m)\kappa + (1-n)(d/2-2)}  \label{IRsolution}
\end{equation}
where $d$ is the space-time dimension and $\kappa$ is one yet undetermined 
parameter. It fulfills some very general inequalities
\cite{Watson:2001yv,Eichhorn:2010zc} which can be summarized as $0.5\le \kappa
<23/38$. With some assumptions on the ghost-gluon vertex its value can
be determined to be $\kappa=0.595$ \cite{Lerche:2002ep}. A further important
property is that there are additional divergences when only some of the momenta 
of the $n$-point functions are vanishing \cite{Alkofer:2008jy}.

Eq.\ (\ref{IRsolution}) especially entails that the ghost propagator and the
three-  and four-point gluon vertex functions are  infrared
divergent. As we will see below this has then profound consequences for the
quark-gluon vertex as well as for  vertices involving 
fundamentally charged matter.

\section{Quark-Gluon Vertex}

Due to the infrared suppression of the gluon propagator, present in the scaling
and in the decoupling solutions, quark confinement cannot be generated by any
type of gluon exchange together with an infrared-bounded quark-gluon vertex. 
To proceed it turns out to be necessary to study the Dyson-Schwinger equation
for the quark propagator together with the one for the quark-gluon
vertex in a self-consistent way \cite{Alkofer:2008tt}. Hereby a drastic 
difference of the quarks as compared to Yang-Mills fields has to be taken 
into account: As they possess a mass, and as D$\chi$SB does occur, the quark
propagator will always approach a constant in the infrared. 

The fully dressed quark-gluon vertex can be expanded in twelve linearly
independent Dirac tensors. Half of the coefficient functions would vanish if
chiral symmetry were realized in the Wigner-Weyl mode. From a solution of the
Dyson-Schwinger equations we infer that  these {\em ``scalar''} structures are,
in the chiral limit, generated non-perturbatively together with the dynamical
quark mass function in a self-consistent fashion. Thus dynamical chiral symmetry
breaking manifests itself not only in the propagator but also in the quark-gluon
vertex.

From an infrared analysis one obtains an infrared divergent solution for the
quark-gluon vertex such that  Dirac vector and {``scalar''} components of
this  vertex are infrared divergent with exponent $-\kappa - \frac 1 2$ if
either all momenta or the gluon momentum vanish
\cite{Alkofer:2008tt}. A numerical solution of a truncated set of
Dyson-Schwinger equations confirms this infrared behaviour. 
The essential components to obtain this solution are the scalar Dirac
amplitudes of the quark-gluon vertex and the scalar part of the quark
propagator. Both are only present when chiral symmetry is broken, either
explicitely or dynamically.

To investigate how this self-consistent quark propagator and quark-gluon
vertex solution relates to quark confinement the anomalous
infrared exponent of the four-quark function is determined.
The static quark potential can be obtained
from this four-quark one-particle irreducible Green function, which
behaves like $(p^2)^{-2}$ for $p^2\to0$ due to the infrared enhancement
of the quark-gluon vertex for vanishing gluon momentum.
Using the well-known relation for a function $F\propto (p^2)^{-2}$
one gets
\begin{equation}
V({\bf r}) = \int \frac{d^3p}{(2\pi)^3}  F(p^0=0,{\bf p})  e^{i {\bf p r}}
\ \ \sim \ \ |{\bf r} |
\end{equation}
for the static quark-antiquark potential $V({\bf r})$.
Therefore the infrared divergence of the quark-gluon vertex, as found in the
scaling solution of the coupled system of Dyson-Schwinger equations, the vertex
overcompensates the infared suppression of the gluon propagator such that one
obtains a linearly rising potential.

\section{Dependence on Lorentz transformation properties}

The above described mechanism which  directly links chiral symmetry breaking
with quark confinement  raises the question about the role of the Dirac
structure in quark confinement. As one  expects that fundamental
charges are confined by a linear potential a next logical step is to
investigate the infrared behaviour of the propagator and the vertex of a
fundamentally charged scalar \cite{Fister:2010yw,Fister:2010ah}.

In contrast to quark Green's functions the tensor structure of the scalar ones
is strongly simplified. Compared to two components in the fermionic propagator,
the scalar propagator features only a single structure. Similarly the vertex
depending on two independent momenta can be decomposed into two tensors (instead
of twelve).

However, a scalar theory has renormalizable self-inter\-actions and therefore the
number of terms in the Dyson-Schwinger and Functional  Renormalization Group
equations are significantly increased. (NB: For the derivation of the
Dyson-Schwinger equations one may employ the MATHEMATICA package DoDSE 
\cite{Alkofer:2008nt}. A package for Functional Renormalization
Group equations will be published \cite{Huber:2011}.)
First, the uniform scaling limit is studied. Applying the additional
constraints on the infrared exponents that arise from the comparison of the
inequivalent towers of Dyson-Schwinger and  Functional Renormalization Group
equations \cite{Fischer:2009tn,Huber:2009wh}, the system of equations for the
anomalous exponents simplifies. One finds the scaling and the decoupling
solutions with an unaltered Yang-Mills sector. In the case of the scaling 
solution for
a massive scalar the scalar-gluon vertex can show either of two behaviors 
 \cite{Fister:2010yw,Fister:2010ah}. 
In one case, which will be discussed here further, it exhibits the
same infrared exponent as the quark-gluon vertex.

However, the uniform scaling displays only part of the potential infrared
enhancements. Vertex functions may also  become divergent when only a subset of
the external momenta vanish. Such kinematic divergences provide a mechanism for
the long-range interaction of quarks as described in the section above. To this 
end it is gratifying to realize that the kinematic divergences of the
scalar-gluon vertex are identical to those of the quark-gluon vertex. These
singularities induce a confining interaction in the four-scalar vertex function as
they did in the case of the four-quark vertex function in the case of scalar
QCD. Corresponding infrared leading diagrams are shown in Fig.~1. Their Fourier
transform according to eq.\ (\ref{IRsolution}) leads to a linearly rising
static potential.

This result provides the possibility that within functional
approaches static confinement is an universal property
of the gauge sector even though it is formally represented
in the functional equations of the matter sector.

\begin{figure}
  \includegraphics[width=.48\textwidth]{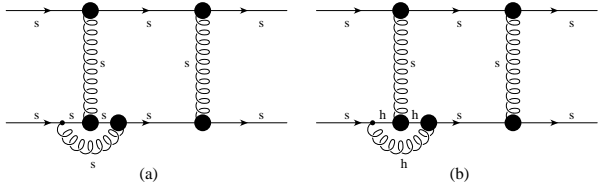}
  \caption{Infrared leading diagrams for the four-scalar vertex function in the
  uniform scaling limit (a) and displaying a kinematic divergence (b).}
\end{figure}

\section{Dependence on the representation / group}

As adjoint charges will not be subject to a linearly rising potential, and as
the deviations from the scaling laws are only possible in case of cancellations
we have investigated whether such cancellations for different representations
and groups occur \cite{Macher:2010}.

Hereby the main difference of course is the color prefactor in the vertex
functions:  A matter-gluon-vertex of a fundamentally charged field is
proportional to a Gell-Mann matrix whereas for an adjoint charge the structure
constant appears. Correspondingly, the color algebra changes.  {\it E.g.\/} the
so-called sword-fish diagram is proportional to the Gell-Mann matrix  for
fundamental scalars but vanishes (due to the antisymmetry of the structure
constants) for an adjoint scalar. Thus, for some would-be infrared leading 
diagrams  there are vanishing color prefactors in the adjoint
representation.

Noting that the exceptional Lie group $G_2$ has a trivial center one expects for
this group the absence of an (asymptotically) linearly rising potential
 already for the
fundamental representation. And as a matter of fact, we find some systematic
cancelations and thus vanishing diagrams for the fundamental representation of
$G_2$. In the comparison of $SU(N)$ and $G_2$ this occurs for infrared leading 
and non-leading diagrams. 

In addition, there are significant differences when comparing the unquenched to
the quenched scalar-gluon vertex Dyson-Schwinger equation.

\section{Summary}

Based on the scaling solution of Functional Equations we have pointed out the
possibility that confinement of fundamental charges is related to an 
infrared divergent matter-gluon vertex. This in turn leads then to a $1/k^4$
singularity in the four-point function and thus to a linearly rising static
potential. In addition, 
for the adjoint representation
and for the gauge group $G_2$ there are systematic cancelations. Whether these
will influence the conclusion on the static potential still has to be
investigated.

\begin{theacknowledgments}
RA thanks the organizers for their outstanding achievement which made this
extraordinarily interesting conference possible.

RA acknlowledges support by the  FWF under grant number P20592-N16,
LF by the ExtreMe Matter Institute EMMI, and
AM by the FWF under grant number M1099-N16.
\end{theacknowledgments}

\bibliographystyle{aipproc}

\end{document}